# Dust from collisions: A way to probe the composition of exo-planets?

*Corresponding Author:*

Andreas Morlok[a,b*]   (morlokan@uni-muenster.de; Phone: ++49 251 83 39069)

Andrew B. Mason[c]   andmas@utu.fi

Mahesh Anand[a,b]   m.anand@open.ac.uk

Carey Lisse[d]   carey.lisse@jhuapl.edu

Emma S. Bullock[e]   bullocke@si.edu

Monica Grady[a,b]   m.m.grady@open.ac.uk

[a]Department of Earth Sciences, The Natural History Museum, London, SW7 5BD, UK

[b]Planetary and Space Sciences, The Open University, Walton Hall, Milton Keynes MK7 6AA UK

[c]Finnish Centre for Astronomy with ESO (FINCA), University of Turku, Tuorla Observatory, Väisäläntie 20, FI-21500 Piikkiö, Finland

[d]Johns Hopkins University -APL, 11100 Johns Hopkins Road, Laurel, MD 20723, USA

[e]Department of Mineral Sciences, National Museum of Natural History, Smithsonian Institution, Washington DC 20560, USA

*Current Address: Institut für Planetologie, Wilhelm-Klemm-Strasse, 48149 Münster, Germany

**Abstract**

In order to link infrared observations of dust formed during planet formation in debris disks to mid-infrared spectroscopic data of planetary materials from differentiated terrestrial and asteroidal bodies, we obtained absorption spectra of a representative suite of terrestrial crustal and mantle materials, and of typical Martian meteorites.

A series of debris disk spectra characterized by a strong feature in the 9.0-9.5 µm range (HD23514, HD15407a, HD172555 and HD165014), is comparable to materials that underwent shock, collision or high temperature events. These are amorphous materials such as tektites, $SiO_2$-glass, obsidian, and highly shocked shergottites as well as inclusions from mesosiderites (Group A).

A second group (BD+20307, Beta Pictoris, HD145263, ID8, HD113766, HD69830, P1121, and Eta Corvi) have strong pyroxene and olivine bands in the 9-12 µm range and is very similar to ultramafic rocks (e.g. harzburgite, dunite)(Group B).

This could indicate the occurrence of differentiated materials similar to those in our Solar System in these other systems.

However, mixing of projectile and target material, as well as that of crustal and mantle material has to be taken into account in large scale events like hit-and-run and giant collisions or even large-scale planetary impacts. This could explain the olivine-dominated dust of group B.

The crustal-type material of group A would possibly require the stripping of upper layers by grazing-style hit-and run encounters or high energy events like evaporation/condensation in giant collisions. In tidal disruptions or the involvement of predominantly icy/water bodies the resulting mineral dust would originate mainly in one of the involved planetesimals. This could allow attributing the observed composition to a specific body (such as e.g. Eta Corvi).

# 1. Introduction

One of the most significant recent developments in the field of infrared astronomy is the ability to obtain spectral information relating to the mineralogical composition of dust involved in exo-planet formation. Similar to our own Earth, exo-planets are thought to form in a circumstellar gas and dust disk around a young star. During the protoplanetary phase, the disk consists of pristine, anhydrous dust, gas, and ices, while unaccreted material is lost. After a short transitional stage, the debris disk stage follows after about 10 Myr. At this stage, bodies grow by collision and accretion into larger planetesimals, and finally into planets (Apai and Lauretta, 2010).

Most debris disks are characterized by Kuiper-belt type cold dust without detectable material in the inner disk. However, a sub-group of 1-2% of the debris disks exhibit infrared emission in the mid-infrared region. This indicates warm dust from collisions between planetesimals in asteroid belt-like regions, where small bodies are consistently ground to dust in a steady-state process system. However, there are also transient events in the form of collisions producing large amounts of dust in short time (Tab.1)(e.g. Löhne et al., 2012; Kennedy and Wyatt, 2013).

While only a few bodies probably reach the size of terrestrial planets, such collisions in the late stage of accretion produce vast amounts of dust comparable to the masses of large planetesimals or planets (Tab.1). The collisions cover the continuum of encounters between similar sized bodies, or, in the case of smaller projectile, large planetary impacts. (Asphaug et al., 2006 and 2010; Lisse et al., 2012; Olofsson et al., 2012).

In this study we focus on collisions between larger planetary bodies as the sources of dust by looking for materials similar to that of differentiated, terrestrial planets. Planetary differentiation was already under way in the debris disk phase of our Solar System, so comparable processes could be expected in other debris disk systems (Elkins-Tanton et al. (2008).

While these are only very rare events (Kennedy and Wyatt, 2013), the identification of planetary materials in such events would help to connect the evolution of these systems to our Solar System.

There is evidence for giant collisions during planetary growth in our Solar System such as the formation of the Moon in the collision of proto-Earth with another planetary body (e.g., Cuk and Stewart, 2012), the loss of crust and mantle of proto-Mercury in a comparable event (Benz et al., 2007), and possibly the Martian hemispheric dichotomy (Nimmo et al., 2008).

Infrared spectra of the dust produced in these processes, taking place in currently forming exoplanetary systems (e.g. Chen et al., 2006) could provide mineralogical and petrological information about the upper layers on the bodies forming there. In the case of debris disks, dust may have been produced by small numbers of colliding planetesimals, possibly even just pairs. A series of studies about dust compositions with the specific aim to characterize these exoplanets or bodies were made by Lisse et al. (2007; 2008; 2009; 2012) and Morlok et al. (2012), who linked differentiated planetary materials to specific debris disks.

Posch et al. (2007), Morlok et al. (2008 and 2011) and Currie et al. (2011) also investigated carbonaceous chondrite type materials in circumstellar environments. A complementary technique is to analyze the elemental composition of planetesimals in old systems in the final evolutionary state. This is done by measuring the heavy element 'pollution' in the H/He-rich stellar atmospheres of white dwarf stars with optical and ultraviolet spectroscopy, where materials of disrupted bodies accreted and settled onto the star (Zuckerman et al., 2007; Jura, 2008 and 2013; Debes et al., 2012; Gänsicke et al., 2012).

For the identification of mineral phases or specific planetary materials in the dust, we need laboratory absorption infrared data for comparison. A number of absorbance/transmission studies have been carried out on specific minerals in recent years (e.g. Jäger et al., 1998, Hofmeister and Bowey, 2006; Bowey et al., 2007; Pitman et al., 2010; Speck et al., 2011). Transmission/absorption studies of bulk planetary materials are, however, relatively rare. Sandford (1984; 1993; 2010) pioneered this field with Morlok et al. (2008; 2010; 2012) and Posch et al. (2007) carrying out additional studies.

Why study bulk planetary materials instead of calculating the compositions based on pure mineral spectra? Pure phases provide an 'idealized' spectrum of a mineral under ideal conditions. Using bulk, natural samples we obtain spectra of the 'real thing', materials that underwent (probably) similar physical events to the materials observed in the astronomical spectra.

A point to consider when studying disk evolution and dust formation is the timing of crust formation. The time needed for solidification and cooling of the magma ocean on a Mars or Earth-like planet is 5 to 10 Myr (Elkins-Tanton et al., 2008). For the Moon, solidification is also possible in ~ 10 Myr, with most of the initial magma ocean solidified in the 1000-year range (Elkins-Tanton et al., 2011). However, there could be differences in the composition of an earlier crust compared to the present day. Many of the observed extrasolar systems used in this study fall into the range from 10 - 20 Myr (Table 1). Systems such as Eta Corvi, BD+20307, HD69830 and HD15407a, on the other hand, may have ages in the billion year range and thus have well developed planetary structures (Table 1).

The aims of this paper are: (a) to present mid-infrared dust absorbance spectra for a representative suite of rocks from highly differentiated, terrestrial-type planetary bodies, (b) to identify characteristic spectral features for these materials that allow the identification in astronomical infrared observations (c) compare the spectra with astronomical observations of dust in debris disks. The identification of such material is of high interest since larger, highly differentiated bodies are the most probable candidates for terrestrial-type, habitable worlds.

**2. Materials and Methods**

**2.1. Sample Selection**

The Martian meteorites consist of three main groups of mafic to ultramafic rocks – shergottites, nahklites, and chassignites (Table 2), with a possible new fourth type just recently discovered (Agee et al., 2013). We have analyzed samples of all three main groups, so the spectra presented are representative of typical Martian materials (McSween Jr., 2007).

Given the size of the impact and events we discuss in this contribution, where the dust masses are at a scale of at least asteroidal bodies (Tables 1, 3), only the most common and abundant materials will be identifiable in the ejecta. Therefore, the sample set for Earth includes granite representing the felsic end of the continental crust, and gabbro, basalt and anorthosite representing the mafic end of the continental and oceanic crust. The suite is complemented by several peridotites representing the upper mantle material. Tektites are representative samples of glassy impact products but also serve as samples of the average upper crust (Kring, 2005). Additional spectra of volcanic glass obsidian and synthetic $SiO_2$ glass to serve as analogues to highly differentiated continental crust were provided by Angela Speck, University of Missouri (Speck et al., 2011). All other samples analyzed in this study are from the collections of the Natural History Museum in London. Sources for these samples and their petrological details are provided in Table 2.

**2.2. Infrared Spectroscopy**

The spectra presented in this study are based on bulk samples and so they represent the whole of a given material, and not specific components or only single minerals.

To avoid any sampling bias due to small sample sizes, we took material from larger (at least 500 mg) powdered representative samples of Martian meteorites.

Since the mid-infrared spectra in the observed dust originate from grains smaller than ~6 µm, the material has to be ground to a comparable grain size for laboratory studies (Min and Flynn, 2010).

The mixture was ground in an agate mortar, and afterwards pressed into a pellet at 10 tons/cm$^2$. We mixed 2-3 mg of powdered sample with 300 mg of powdered potassium bromide (KBr), using a well-tested preparation technique for grain sizes with an average of about 1 µm (Chihara et al., 2002; Koike et al., 2003; Imai et al., 2009).

To avoid bands due to absorbed water, the pellets were dried in an oven at 100 °C for three days.

Different preparation methods used to grind bulk materials may result in differences in band positions and shape. The differences are minor however, and hand-grinding (this study) or ball-mills usually provide comparable spectra in the mid-infrared. For a detailed discussion of the effects of preparation on spectra features see Imai et al. (2009) and Lindsay et al. (2013).

The analyses were carried out at the Natural History Museum in London, using a Perkin Elmer Spectrum One workbench. From each pellet, a spectrum was obtained from 2.5 to 40 µm, with a spectral resolution of 4 cm$^{-1}$, adding 50 scans for each sample.

Positions of characteristic features were obtained using Origin Pro 8. Feature positions have an instrumental uncertainty of 0.05 µm (Morlok et al., 2006). In comparisons between laboratory data and astronomical spectra, band positions are regarded as similar in the 8-13 µm range when they fall within 0.25 µm. This is necessary to allow for variations due to differences in composition or analytical accuracy (see discussion in Morlok et al., 2005).

The results are presented in terms of absorbance (A) and normalized to the same intensity. The results are presented in Figure 1a and b and in comparison with astronomical spectra in Figure 2 and 3.

## 2.3. Astronomical Spectra

For a discussion of the selection criteria of the astronomical spectra see Section 4.

The spectrum of Beta Pictoris was provided by Y.K. Okamoto (Ibaraki University) and obtained using the Cooled Mid-Infrared Camera and Spectrometer (COMICS) at the Subaru Telescope in the 8-13 µm range with a spectral resolution of R=(λ/D λ) < 250 (for more details see Okamoto et al., 2004).

The infrared spectra of BD+20307, HD165014, HD15407a, HD23514, HD172555, ID8, HD145263, P1211, HD113766, Eta Corvi and HD69830 were made using the InfraRed Spectrograph (IRS) on board of the Spitzer Space Telescope. The spectral range was 5–35 µm (except for HD69830 with 7-35 µm), and the spectral resolution R=90-100 in the 5-14 µm region, the part of major interest in this study.

For sources and further details please see Gorlova et al., 2004; Beichman et al., 2005; Chen et al., 2006; Lisse et al., 2007, 2008, 2009 and 2012; Rhee et al., 2008, Melis et al., 2010; Fujiwara et al., 2010, 2012; Weinberger et al., 2011; Meng et al., 2012.

The astronomical spectra in Figure 2 are in Flux (Jy), those in Figure 3 and 4 Flux spectra normalized to the continuum: $(F\nu)/(B\nu(T))$.

In order to quantify these observations, we calculated intensity ratios of the integrated areas from 9.0-9.6 µm (Band A) and 10.8-11.4 (Band B) (Fig.3a-c). These areas show characteristic features in both laboratory and astronomical spectra (see 4.1.4 for more details). To allow a direct comparison between the two sets of data, the spectra had to be normalized.

This was carried out by first removing the continuum in the 8-13 µm regions from several selected spectra by dividing the flux ($F_\nu$) with a blackbody continuum ($B_\nu(T)$). This was calculated with a Planck function for the dust temperature T and distance of the debris dust r. Parameters used to process the astronomical data were T=560K and r=0.7-4.4 AU for HD165014 (Fujiwara et al., 2010), T=335K and r= 5.8 AU for HD172555 (Lisse et al., 2009), T=420K and 1 AU for HD23514 (Meng et al., 2012), T=520 K and r=0.6 AU for HD 15407A (Fujiwara et al., 2012), T=864 K and r=0.5 AU for ID8 (Oloffsson et al., 2012), T=800K and r=<3.2 AU for Beta Pictoris (Okamoto et al., 2004), T=490K and r=1.8 AU for HD113766 (Lisse et al., 2008), T=390 K and r=3 AU for Eta Corvi (Lisse et al., 2012).

For a direct comparison between the laboratory absorbance data with the resulting astronomical emissivity spectra, it was necessary to normalize the spectra to a baseline. The baseline was calculated between the low-points of the processed spectra in the 5-20 µm regions, which should indicate the lowest intensity of the silicate signal in this part of the spectrum. Finally, all spectra used in the comparison were normalized to 1 based on the strongest spectral feature in the 8-13 µm regions. The focus of this investigation is on the ratio of silica and silicates, so the omission of underlying signal from other mineral species by the baseline procedure will not affect the results significantly.

## 3. Results

The characteristic band positions of the samples are presented in Table 4a (Martian) and 4b (Terrestrial), and the associated spectra are shown in Fig.1a and 1b. Details of the mineralogical composition are provided in Table 2. Band positions for the typical mineral components are presented in Table 5.

### 3.1 Shergottites

Shergottites analyzed in this study (Fig.1a, Tables 2 and 4a) can be separated into two subgroups that are easily distinguishable by their bulk spectra. The basaltic pyroxene-phyric meteorites Zagami and Shergotty have spectra dominated by typical clinopyroxene features at 9.4, 10.4, 11.3, and 19.9 μm (Koike et al., 2000; Chihara et al., 2002).

Few plagioclase features are recognizable in the spectra (e.g. at 10.65 μm), since most of the feldspar has been transformed to amorphous maskelynite in shergottites as result of shock from impact (e.g. McSween and Treiman, 1998). Therefore the intensities of the crystalline features in the 10 μm range may be affected by a 'continuum' of the maskelynite spectrum (Palomba et al., 2006; Gyollai et al., 2010; Johnson, 2012).

The basaltic shergottite Los Angeles is probably affected most by these shock effects (Walton and Spray, 2003). The spectral features are generally broad and rounded; the features in the 8 – 13 μm are very weak; we see a strong single feature at 9.52 μm, with shoulders, while the strong band at ~21 μm is clearly shifted in comparison to the equivalent feature at 19.9 μm in Shergotty and Zagami. Our data are in agreement with the spectra obtained by Sandford (1984).

The second group of shergottites is the olivine-phyric basalts represented in our study by Dar al Gani 476 and SaU 005 (Fig.1a, Table 2and 4a). Again, the two samples show very similar spectra, demonstrating a large degree of intra-group homogeneity. The strongest features are olivine bands at ~11.3 μm (Morlok et al., 2006; Pitman et al., 2010) overlapping with a strong Ca-pyroxene band at the same position. Most of the other features are typical of clinopyroxene (Koike et al., 2000). A strong feature at 7 μm could be ascribed to carbonate (Koike et al., 2000). This mineral group occurs in very small amounts in Martian meteorites in cracks and veins as product of aqueous alteration, but could also be the result of terrestrial alteration, since the samples are desert finds (e.g. Hallis et al., 2012).

### 3.2 Nakhlites

The band positions in these clinopyroxenites (Lafayette, Governador Valadares, Y 000593, Nakhla, and MIL 03346) are nearly identical (Table 4a; Fig.1a). The only significant difference between the spectra is the relative intensity of the three strong features between 8 and 12 μm. Characteristic features of Ca-pyroxenes are at 9.4, 10.3, 11.4, and 20.9 μm (Koike et al., 2000). Here the feature at ~20 μm also splits into two minor peaks, in contrast to the shergotittes. Earlier analyses of Nakhla and Lafayette are comparable to those in our study (Sandford, 1984).

The variable intensity of the 11.3 μm band could be due to the strong olivine feature at the same position (Morlok et al., 2006), thus reflecting the variable olivine/pyroxene ratio in the nakhlites. Alternatively, varying contents of plagioclase/maskelynite could interfere with the features at ~10 μm.

### 3.3 Chassignites

The Martian dunite Chassigny, as expected, shows only typical olivine features (Fig.1a, Table 4a). These are major characteristic bands at 10.29, 11.3, 20.24, 24.81, ~29, and ~35 μm. These bands indicate an intermediate to forsteritic composition (Koike et al., 2003; Morlok et al., 2006).

**3.4 Terrestrial**

The granite spectrum is dominated (Fig.1b, Table 4b) by plagioclase and alkali feldspar features, with the most characteristic bands at 8.8, 9.6-10, 17, and 23.4 μm (Suhner, 1986; ASTER Spectral Library http://speclib.jpl.nasa.gov/documents/jhu_desc). The strongest quartz bands are at 9.15, 12.8-12.9, and 21.6 μm (Farmer et al., 1974; Koike et al., 1994). Obsidian, a silica-rich glass, has its strongest band at 9.3 µm and 12.2 µm (Speck et al, 2011). The tektites (Australite and Bediasite; Figure 1b; Table 4b) have very similar spectra, which are essentially smoothed versions of the granite spectra with a broad feature at 9.2-9.3, a weaker one at 12.6 µm, and another strong feature at 21.6-21.7 µm. These features are comparable to that of glassy $SiO_2$ material (Fig.1b, Table 4b), which has its main band at 8.97 and 12.17 µm. (Speck et al., 2011).

Feldspar-rich anorthosite (Figure 1b; Table 2 and 4b) has characteristic plagioclase bands at 10 μm, and at 25.5 μm. Other significant features are at 13.5, 17.2, and 18.5 μm (Suhner, 1986; ASTER Spectral Library http://speclib.jpl.nasa.gov/documents/jhu_desc).

The different basalts (Figure 1b; Table 2 and 4b) show similar features, reflecting the high feldspar content (Tab.2): a strong band from 9.8 to 10.1 μm, and a second strong feature between 21.2 and 23.4 μm. Further characteristic bands are between 12.8 and 13.9 μm (Suhner, 1986; ASTER Spectral Library http://speclib.jpl.nasa.gov/documents/jhu_desc; Speck et al., 2011).

The gabbro sample has two very strong features at 10 and 21.8 μm, with further bands at 15.9 and 17 μm (Figure 1b; Table 2 and 4b). Similar to the basalts, the mineralogy can be explained by a mixture of plagioclase features. Only the bands at 9.3 and 12.5-12.8 μm may require a quartz component (Farmer, 1974). A suite of typical mantle peridotite samples (dunite, harzburgite, and pyroxenite) reflects the presence of different minerals and their abundances, mainly olivine and pyroxene (Figure 1b; Table 4b). Main features in dunite and harzburgite are typical forsterite bands e.g. at 11.3, 11.9, ~16.4, 19.8 and ~24 μm (Koike et al., 2003). Significant pyroxene bands are visible at 9.3 (shoulder), ~10.2 and ~10.5 μm (Koike et al., 2000). Our pyroxenite sample shows bands of ortho- and clinopyroxene with characteristic features at 9.31, 10.17, 11.4, and 15.8 μm (Koike et al., 2000; Chihara et al., 2002).

**4. Discussion**

In the discussion, we compare the laboratory spectra of differentiated terrestrial materials with those of dust from warm emission in debris disks. Dust in this type of debris disk could be formed by a range of events, from mutual collisions in asteroid belts, cometary dust, dynamic instabilities that are comparable to the Late Heavy Bombardment in our early Solar System, and debris from giant collisions of planetesimals during the late stage of planet formation (Lisse et al., 2007; 2012; Asphaug et al., 2010; Olofsson et al., 2012).

The materials involved could range from primitive type materials from undifferentiated asteroids or comets, to achondritic rocks from differentiated smaller planetesimals to crustal and mantle materials from differentiated (proto) planets (see sources in Table 1).

Similarity between spectra alone will give us only limited information about the processes in a specific debris disk system. The events in a debris disk can involve planet wide changes, where both/several bodies are disrupted, down to smaller scale surface event such as planetary impacts (e.g. Asphaug, et al., 2006). This makes a discussion of the various dust creating collision events and the involved dust processing necessary, in order to see if there are any characteristics in these events that can be related to the dust properties and help to clearly link them to planetesimal collisions (or not).

Below we will discuss the various scenarios for the dust production, the involved dust masses and the physical state of the material, too (4.2. Dust Sources).

The focus of this study is on the search for material from terrestrial planets. One selection criteria for astronomical mid-infrared spectra were high dust masses comparable to or higher than those of the asteroid belt in our Solar System (Tab.1 and 3). Dust masses were not available for all systems, and the masses are also highly model dependent (see 4.2.1. Dust Masses). A second criterion was that planetesimal collisions were suggested as dust sources for these systems in earlier studies (see sources in Tab.1). For this reason, we also included a couple of systems with lower masses (HD69830, Beta Pictoris, see Tab.1).

## 4.1. Grouping and Comparison

### 4.1.1. Grouping of Terrestrial and Martian Spectra

We divide the Martian spectra into three general groups based on their spectral features. Most nakhlites and shergottites show a series of pyroxene features in the 9-12 μm region. Chassignites, a further group, have characteristic olivine bands. A third group is comprised of the highly shocked shergottite Los Angeles, with smooth features and a strong band at 9.5 μm (Figure 1a, Table 4a).

Terrestrial samples can also be divided into three main groups. The crustal materials (granites, basalts, anorthosite, and gabbro) show a strong feature at ~10 μm, and another one at ~22 μm (Figure 1b, Table 4b). Tektites are different with their main silica band at 9.2-9.3, and a generally 'smoother' spectrum with broader bands such as the Los Angeles sample. The suite of mantle materials (dunite, harzburgite, and orthopyroxenite) is significantly different; they mainly show olivine, and pyroxene features.

### 4.1.2 Grouping of Astronomical Spectra

As with the laboratory spectra, the first step is to determine if the astronomical mid-infrared spectra also show similarities which allow them to be grouped. This would make comparisons easier, and, more importantly, could point toward potential systematics among the observed forming bodies. Comparable astronomical spectra could mean distinct classes of bodies (such as differentiated and primitive, undifferentiated bodies) or even specific components (parts of mantle, crust) of larger bodies involved. Most of the obtained debris disk spectra are easily sorted into two groups based on their characteristic spectral features.

Group A (Table 6; Fig.2a) consists of HD23514, HD15407a, HD172555 and HD165014. The first two systems have their main band at 9.0-9.1 µm and 20-21 µm, the other two at 9.3 µm. HD165014 also a distinct enstatite feature at 10.4 μm.

Group B (Table 6; Fig.2b) consists of BD+20307, Beta Pictoris, HD145263, ID8, HD113766, HD69830, P1121, and Eta Corvi (Okamoto et al., 2004; Lisse et al., 2007, 2008, 2012; Melis et al., 2010; Weinberger et al., 2011; Meng et al., 2012). Characteristic of this group are two bands at 9.9-10.3 µm and 10.8-11.3 µm with varying intensity. BD+20307 only shows a weak feature at 10.9 µm, while the band is strongest in HD69830 (also 10.9 µm) and Eta Corvi (11.3 µm). The spectrum shows more olivine features at ~ 19 µm, and a strong, broad band at 23-24 µm (Koike et al., 2003; Morlok et al., 2006).

**4.1.3. Comparison of Laboratory and Astronomical Spectra**

Band positions in group A overlap with glassy materials of crustal composition such as tektites and obsidian with main bands at 9.0-9.3 µm. Synthetic $SiO_2$ glass with the main band at 9 µm could confirm high silica content, pointing towards highly differentiated, crustal material. Inclusions from the mesosiderite Emery from a study of achondrites (Morlok et al., 2012) are debris from asteroidal or planetesimal crust mixed with core metal in a collision (Scott et al., 2001; Morlok et al., 2012). The Inclusions are also similar with a feature at 9.1 µm.

The bands in the 18-21 µm regions for group A differ slightly from the comparable laboratory spectra with 21.2-21.7 µm. This could be explained by admixture of mafic mantle material, which has a strong band at higher wavelengths (see below).

HD15407A, HD23514 and HD172555 have features between 12.4 and 12.6 µm, overlapping with a band at 12.5 µm in tektites and Emery (Fig.2a; Tab.4b and 6).

Shergottite Los Angeles 001 has a main band at 9.5 µm and is also similar to the range of group A, although the feature at 21 µm differs from those of HD172555 and HD165014, but is close to HD23514. The other Martian samples of shergottites and nakhlites as well as crystalline terrestrial samples are less comparable to group A.

Earlier studies (Table 1) of the detailed mineralogy of the debris disks showed high $SiO_2$ (silica), feldspar and pyroxene for various members in group A. This supports the findings of highly differentiated surface materials (Rhee et al., 2008, Lisse et al., 2009; Fujiwara et al., 2010, 2012). Amorphous material (pyroxene and/or silica) was suggested for HD15407A and HD165014, HD172555 (Lisse et al., 2009; Fujiwara et al., 2010 and 2012). For HD165014 also a differentiated source (similar to aubrite achondrites) was suggested, formed in a collisional cascade of asteroids (Fujiwara et al., 2012).

Group B is comparable in band positions with the suite of mantle materials (peridotites), chassignites, in particular for bands of dunites and harzburgites at 10.1 - 10.3 µm, 11.3-11.4 µm, but the ranges overlap barely. The olivine-rich brachinites included in an earlier study of achondrites also fall into group B (Morlok et al., 2012).
Lisse et al. (2008) and Morlok et al. (2012) previously identified the source of dust in HD113766A as a differentiated, olivine-rich ureilite-type body. In general, earlier studies showed an olivine-dominated mineralogy with additional crystalline/amorphous pyroxene for most systems in group B (Table 6).
Undifferentiated, chondritic parent bodies as a potential dust source have also been suggested for HD69830, HD113766A, and HD145263 (Olofsson et al., 2012; Lisse et al., 2007; Honda et al., 2004). So olivine-rich dust would make it difficult to distinguish in Group B between mafic mantle material from a differentiated planetesimal and dust from chondritic, primitive bodies.

Not all warm debris disk spectra fall in into group A or B. EF Cha shows mainly phyllosilicate features with a strong main band at 10 µm (Currie et al., 2011). A Spitzer spectrum of the whole Beta Pictoris system (Chen et al., 2007) differs from the inner disk spectrum of the same system presented here (Okamoto et al., 2004). It shows many features in the 10 µm region, possibly due to a mixture of various components. Systems such as HD98800 show only a feature of amorphous silicates at 10 µm (e.g. Olofsson et al., 2012). These dust compositions also could point towards more primitive, chondrite-type compositions.

**4.1.4. Band Ratios**

In a scenario involving collisions of differentiated planetesimals as dust source, a strong feature at 9.0-9.5 μm accompanied by a band at > 20 μm could be a potential sign for crustal-type material.

The position of the strongest feature in the 8-13 µm region and the ratios between the intensity of characteristic bands A and B, are used to quantify groups A and B and their connection to the astronomical data (Fig.3a-c). We compare the integrated intensities for Band A (9.0-9.6 µm) and Band B (10.8-11.4 µm), the areas below a 0.6 µm range in order to avoid artefacts due to spurious features or low signal to noise ratio.

Band A covers the characteristic features of the mainly amorphous/heavily shock processed, $SiO_2$/feldspathic/pyroxene-rich group of materials. They represent a crustal/upper mantle component such as tektites, obsidian, glass $SiO_2$, shergottite Los Angeles and asteroidal crustal material from mesosiderite Emery.

Band B is the integrated intensity of the area around the strong olivine feature at 11.2 µm, which is used as a proxy for ultramafic, mantle type materials. There are caveats – pyroxene also has strong features in the area of Band B. Still, the ratio of the two bands should serve as rough proxy for the content of a highly differentiated, upper mantle/crustal-like component. In addition, we use the position of the strongest feature in this area as second indicator, to use the shift of the feature at ~9 µm to separate the various groups (Fig.4) (Brown and Musset, 1993; Lewis, 1995).

Band ratios and positions of the laboratory data and selected astronomical spectra confirm the general picture (Fig.4). The spectra of group A, HD172555, HD15407A, HD23514 and HD165014, plot within an order of magnitude of the tektites, mesosiderite Emery and shergotitte Los Angeles (Fig.4). Obsidian and glassy $SiO_2$ show higher ratios, indicating that the material from the debris disks could contain high amounts of more mafic materials, as would be expected in such large scale events (See 4.2.2. Mixing). However, since highly differentiated crustal material would be expected in a mixture of crustal and mantle, they could be end members in a mixture with more mafic glasses. Crystalline crustal materials (e.g. granites) form a different group owing to the strongest mid-infrared band falling clearly into the ~10 µm area.

Typical Group B member ID8 (Fig.3b-c, Fig.4) is similar to mafic or mantle type materials such as terrestrial harzburgite or dunitic achondrite Brachina. Those with the 11.2 µm band dominating (e.g. Beta Pictoris, Eta Corvi or HD113766) are similar to terrestrial dunites, and the achondrite group of ureilites. The similarities of latter type of material with Eta Corvi confirm the findings in Lisse et al. (2012). Achondrites of the abundant HED group (Howardites, Eucrites, and Diogenites) fall between the two groups.

**4.2. Dust Sources**

Near-perfect accretion where the masses of the projectile and the target merge without substantial mass losses only takes place at low impact/collision speeds (Asphaug, 2010). Depending on the angle of collision and the mass ratio of the involved bodies (m=mass of impactor; M=mass of target), the transition from accretion to substantial mass loss takes place at impact velocities between 1.1 to 1.6 times the escape velocity of the whole mass of the involved bodies. The resulting debris in the form of vapor, melt and dust is potentially the material we observe in debris disks. Some or most of the material produced in such events may be re-accreted later on, while part of it is permanently ejected (compare Chambers, 2013).

Modelling of collisions shows generally comparable outcomes over the whole mass range from planetesimals to super-Earths (with masses of up to 15 times that of the Earth), including water/ice-rich bodies. So an Earth sized body could be as well a target or the projectile in the discussed collisions (Marcus et al., 2009; Asphaug, 2010; Marcus et al., 2010a, b).

Collisions between planetary bodies that could sometimes produce dust material can be divided into several 'regimes'. Giant collisions refer to cases where planetary bodies of similar sizes at the larger end of size distribution collide at velocities comparable to the mutual escape velocity (Asphaug et al., 2006; Asphaug, 2010). Benz et al. (1988 and 2007) modelled the loss of the mantle of proto-Mercury in a collision with a differentiated body at high impact velocities. Most of the ejected material in these models is not re-accreted, and, therefore became a source of dust. Related models for the formation of the Moon also involve the collision of planetary bodies with a mass ratio of down to only m/M≥0.05 (Cuk and Stewart, 2012).

Giant collisions between planetesimals were among the suggested dust sources for most systems used in this study (references see Table 1). Since the various types of collisions possibly affect the dust compositions in different ways, we discuss these points in the following in an attempt to link these scenarios to the possible dust compositions.

Hit and run planetary collisions provide a scenario where two bodies collide, but where the projectile escapes without being accreted entirely into the target, but is instead stripped of parts of its mantle. Such a scenario was proposed for Mercury as the minor body in an encounter (Asphaug et al. 2006, Asphaug, 2010; Stewart and Leinhardt, 2012).

In tidal disruptions the projectile passes the target very closely, but without direct contact. A study of such an encounter of Mars/Moon sized differentiated bodies (m/M=0.1) resulted in the smaller body stripped of parts of the mantle, with the target staying completely intact (Asphaug et al., 2005 and 2006).

In planetary impacts the spherical shape of the target plays a role in impact mechanics, and the impact does not completely reorganize the target, but instead leaves a trace in the form of a crater. In a model of the formation of the Northern Lowlands of Mars (17000 –20000 km diameter) impactors in the size range of 250 - 2830 km ejected material to escape velocity (Marinova et al., 2011).

**4.2.1. Dust Masses**

A clue to the processes involved in the dust formation of a given system is the dust mass. It should be kept in mind that these masses are highly model dependent with an error of about 50% (C.Lisse, pers.comm), thus they only provide a rough estimate.

In several systems, the orders of magnitudes for dust masses fall into a size range between the masses of the bulk asteroid belt in our solar system ($10^{24}$g) and the material comprising the bulk Earth at ($6\times10^{27}$g). These masses are comparable to Mercury and Mars ($10^{26}$g) or Moon ($10^{25}$g), but also of the terrestrial crust ($10^{25}$g) or mantle ($10^{27}$g) (Table 3). This range of dust masses would (in the context of collisional events) point towards giant collisions or Hit and Run events (Lodders and Fegley, 1998; Asphaug et al., 2006; Lisse et al., 2012). Of these, HD172555, HD15407a, and HD23514 are potentially silica-rich systems of group A, and HD113766, Eta Corvi, ID8 of mafic group B (Table 6).

In planetary impacts, we expect lower dust masses. The scenarios for Mars (Marinova et al. (2011) have ejecta with masses of up to $2.7\times10^{25}$ g escaping Mars. This is more than the dust masses of BD+20307, Eta Corvi, areas of high dust density ('lumps') in Beta Pictoris and HD69830 ($10^{21}$-$10^{25}$ g; Tables 1 and 3) (Okamoto et al., 2004; Lisse et al., 2007; Currie et al., 2011; Fujiwara et al, 2012).

However, grinding sequences in an excited asteroid belt can also produce comparable amounts of dust ($10^{16}$ g - $10^{24}$). For example, for HD69830 a carbonaceous-chondrite-type composition based on disrupted asteroids was modelled (Lisse et al., 2007 and 2012).

**4.2.2. Mixing**

An important point is that in a collision the dust usually comes from at least two bodies involved. In the observed dust in debris disks, possibly several planetesimals in grinding sequences can be involved. In so far the models used in this study with only two colliding bodies are possibly simplified.

In Hit-and Run events, the debris is mostly from the projectile, but a component of the target has to be expected when the materials of the mantles mix during the contact (Asphaug, 2010). Models for the ejected material in lunar formation/giant collision scenarios show compositions ranging from ~90% projectile to over 90% material of the target in models using a fast-spinning Earth providing higher kinetic energy (Asphaug et al., 2006; Canup 2004; 2008, 2012; Cuk and Stewart, 2012; Reuffer et al., 2012).

In models of mass ejecta from terrestrial planets in impacts, high impact velocities result in several times the impactor's mass ejected to escape velocity: about three times from a Mars type body (Marinova et al. 2011), and 4.2(asteroid) to 12 (for cometary impacts) times the impactors mass for lunar impacts (Shuvalov and Artemieva, 2006). This would result in a clear signal of the crustal/mantle material of the target, especially in the case of a cometary impactor, since the high volatile component would additionally reduce the impactors' mineral signature.

Similar, the involvement of icy/wet bodies as projectiles would indicate that we see mainly the composition of one of the bodies, since water and ice have clearly different infrared features compared to minerals (Lisse et al., 2012). Of the presented systems only Eta Corvi shows large amounts of cold material. This is interpreted as ice from the projectile(s), presumably Kuiper-belt like objects crossing into the inner system (Lisse et al., 2012). In case of an impact or collision, this potentially would only show the mineralogy of one of the involved planetesimals.

**4.2.3. Potential scenarios**

Mantle material dominates the silicate part of a terrestrial planet such as Earth by two orders of magnitude (Lodders and Fegley Jr, 1998). So a mixture dominated by mantle material (group B) should usually be expected in planet wide events such as giant and Hit and Run collisions, or in tidal disruptions. The dust masses of most systems could points towards such larger events (see 4.3.1).

The low intensity of features from mantle type minerals (olivine) and the predominance of crustal phases (silica, pyroxene, feldspar) make it difficult to link spectra of group A to large scale collisional events.

Planetary impacts might allow ejection of mainly surface material, but the mafic debris of BD+20307, Beta Pic, Eta Corvi and HD69830, all well below the $10^{25}$g possibly ejected during the formation of the Martian lowlands, are all in mafic group B.

This could also confirm one of the suggested scenarios in Lisse et al. (2012), which featured the impact of several icy planetesimals on a differentiated planetesimal. The 'warm' rocky part of Eta Corvi was identified to be comparable to the ureilites; mantle material from differentiated bodies (Lisse et al., 2012) (Figs. 4).

So what about group A? A grazing-type Hit and Run collision might result in dust dominated by silica-rich surface material. Studies in high speed giant collisions (Asphaug et al., 2005) show the target losing the top layers of its mantle (which would include the crust), without being disrupted. This could also lead to a chemical/mineralogical fractionation with a high ratio of crustal-type Group A material compared to relatively less mantle material of Group B in the dust.

In a study of HD 172555 (Lisse et al., 2009), $SiO_2$ rich debris is explained as an ongoing condensation sequence from a recent collision, where most of the vaporized component may still be in the vapor phase. However, the spectrum exhibits signs of abundant SiO gas, which is lacking in the other silica-rich candidates.

Tidal Disruption could explain the silica-rich group A by having mainly crust/upper mantle of a planetary-sized projectile stripped off in a close (non-contact) encounter with a larger planet. Such an event would also exclude material from a second body, so we could study the composition of a single body (Figure 4, Table 1 and 6).

However, the question of whether it is actually mechanically possible to separate only the upper layers of the planetary mantle must be considered. A potential point for the localization of shear forces is the core/mantle boundary, so separation should mainly be expected there, which would result in the stripping off of the bulk-mantle and the crust (Asphaug, 2010). Shear localization is also observed in chondritic-type material, so various transitional zones in the planetary mantles may also make shear localization possible. For example, a spot analogue to Earth's Mohorovičić discontinuity, which separates crust and mantle, would lead to preferred loss of silica-rich crustal material (van der Bogert et al, 2003; Asphaug, 2010).

Tidal disruption of planetesimals or planets has already been identified as source of dust in the case of metal-pollution of white dwarf stellar atmospheres observed via optical spectroscopy. For example, the high Al and low Fe abundances in white dwarf NLTT 43806 is considered to be a potential signature of a planetary lithosphere, which however is suggested to have been separated in an earlier impact event (Zuckerman et al., 2011). Klein et al. (2010), Gänsecke et al. (2012) and Xu et al. (2013) also suggest differentiated bodies as 'pollutant' for several white dwarf atmospheres.

4.2.4 Dust Creating Processes and Physical State of Material

The grain size of the ejecta is important since the strongest infrared emission comes from grains smaller than ~10 µm (Min and Flynn, 2010). Most of the material that is ejected as vapor and melt in giant collisions is forming particles down to the µm-size range (Benz et al., 2007; Takasawa et al., 2011). Johnson and Melosh (2012) model an average size of droplets condensing from the vapor plume of a Chixulub-type impact of about 250 µm. Larger droplets and dust grains would be ground to particles smaller than 1 µm by collisional fragmentation at relative velocities higher than about 1 km/s. Further processes eroding dust are sputtering by energetic particles or sublimation close to the star (for a review of dust processing in debris disks sees e.g. Mann et al., 2006 and 2014).

The actual dust creating processes and the resulting physical state of the debris are also important to consider when interpreting the results. A central question is if the crystalline material observed in debris disks is pristine. Most of the ejected debris of giant collisions is in melt or vapor state (Asphaug et al., 2006, 2010; Canup 2008, 2012). For example, vaporization of $SiO_2$ begins at pressures of 75 GPa, which corresponds to impact speeds typically achieved in the inner solar system (5 to 40 km s$^{-1}$). For this reason, abundant vapor has to be expected even at relatively low speeds (Kraus et al., 2012). Scenarios using grazing collisions at lower velocity (< 15 km/s) allow for some solid material to survive in the ejecta (Wada et al., 2006), but would also produce vapor.

At least for HD172555 vaporization followed by condensation has been suggested (Lisse et al., 2009). For more details about gas and melt after giant collisions see e.g. Pierazzo et al. (1997) and Visscher and Fegley, Jr. (2013).

However, the disruption of a planetary body in larger fragments followed by a grinding sequence of the remnant larger debris still may allow the survival of crystalline dust (Olofsson et al., 2012). A study of impact heat effects on differentiated planetesimals up to 1000 km in diameter shows difficulties in melting substantial parts of the bodies even in near-disruptive events (Keil et al., 1997). In studies of impacts on planetary bodies, the amount of solid material among the escaping ejecta is found to be of the order of the projectiles mass for lunar impacts (Shuvalov and Artemieva, 2006).

Annealing is a process that would allow amorphous materials from cooled melts to crystallize in the debris disks. Laboratory experiments point towards higher activation energies for enstatite compared with forsterite as source for olivine rich material (e.g. Fabian et al., 2000; Thompson et al., 2002). Annealing studies of amorphous silicates resulted in olivines with iron/magnesium ratios similar to that observed in debris disks (Nuth and Johnson, 2006; Murata et al., 2009). These findings could explain the olivine-rich dust found in group B. Crystallization processes in debris disks are reviewed in Olofsson et al. (2012), vaporization, condensation and crystallization processes for silicates in circumstellar environments in Mysen and Kushiro (1988).

There are many uncertainties as to what degree the processes discussed above affect the physical state of the debris dust. However, the observed crystalline features in the produced dust are possibly not pristine, but from recrystallized material. This has to be kept in mind for direct comparisons with laboratory analyses of crystalline planetary materials. Laboratory spectra of glass on the other hand, are less affected by this problem, since they could provide a direct analog to shock melted and rapidly cooled material.

**5. Conclusions**

The comparison of mid-infrared spectra of terrestrial or Martian rocks with astronomical spectra of debris disks allow to identify two major groups with similar spectral characteristics.

Group A consists of HD23514, HD15407a, HD172555 and HD165014 and is comparable to laboratory spectra from terrestrial and other differentiated bodies that underwent high temperature and/or collisional events: terrestrial tektites, obsidian, $SiO_2$ (representing terrestrial crustal material), Los Angeles (a highly shocked Martian surface sample), and silicate-rich inclusions from mesosiderites (crustal debris from planetesimal collisions) (Scott et al., 2001).

Group B (BD+20307, HD113766, HD69830, HD145263, Eta Corvi, Beta Pictoris, P1121, ID8) shows features at ~10 and ~11 µm. These spectra are comparable to mafic and dunitic mantle materials such as Martian meteorite Chassigny, terrestrial peridotites and also asteroidal brachinites.

This similarity between differentiated planetary materials and debris disk dust could indicate that materials in terrestrial planet formation between our solar system and extrasolar systems are comparable. In the context of collisions between large differentiated planetesimals, groups A and B could be composed of the silicate portions of a terrestrial planet, crust and mantle.

High dust masses and predominantly mafic materials of group B could be a result of mantle and minor crustal materials mixed in giant or Hit and Run collisions. Hit and Run-type collisions could also provide an answer to the 'missing mantle paradox', the relative lack of olivine-dominated mantle type material among achondrites (Burbine et al., 1996; Asphaug et al., 2006): the olivine-rich dust observed in these debris disks could be the dust produced by stripped mantle material.

For type A dust to be produced, high energy processes such as evaporation and condensation or grazing-type Hit and Run collisions could be candidates (Asphaug et al., 2006; Lisse et al., 2009).

Group A spectra all fall into the high dust mass group ($10^{26}$g). While calculating the dust masses is highly model dependent, this could indicate that we mostly see such large scale impacts.

Tidal disruption of projectiles provides a possible scenario where we would observe the composition of a single body, without mixture of materials from target and projectile. In large impacts it is possible that the majority of the escaping material is from the target (Marinova et al., 2011), so the dust composition in question could actually be the chemical/mineralogical signature not just of a 'bulk' planetary mantle, but that of a relatively regional geological structure on another planet.

An icy body as impactor will not 'contaminate' the spectral rock features of the second planetesimal (Marcus et al., 2010a, b; Reufer, 2012). In case of a collision, this potentially makes Eta Corvi a candidate for dust that came primarily from the mantle of an exo-planet.

For group A the usually high dust masses and silica composition make a potential, but not definite case for the disruption of larger planetesimals. For group B, the situation is more difficult, because there are alternative dust sources such as chondritic bodies. This illustrates the difficulties in distinguishing between materials with similar mafic mineralogy but entirely different origins. While the formation of the dust in the discussed debris disks is a possibility based on the composition, it is still at the moment only one potential scenario. Further details about the system, such as potential planets and their orbits are possibly needed to clearly distinguish dust from different events.

Suggested future work to clarify these issues include more detailed modelling of the fate of the ejecta – how much actually ends as non-accreted dust (Asphaug et al., 2010), and how much stays in orbit for a statistically significant amount of time. Also, variations in composition and structure of terrestrial planets should be included in models of collisions; for this the ratios of group A and B materials in debris disk spectra could be compared.

On the analytical side, changes of spectroscopic properties during dust formation and after ejection would be useful. Here the effects of the space environment (space weathering) on the dust would be of interest. Also experiments on the crystallization of melt droplets and the formation of crystalline phases would help to characterize the dust.

**Acknowledgements:**

Many thanks to Angela Speck (University of Missouri) for providing the spectra of obsidian and $SiO_2$. We would like to thank the Meteorite Working Group (MWG), National Institute of Polar Research (NIPR), for providing Martian samples.

Many thanks to Caroline Smith, Dave Smith for helping with the Martian and terrestrial rock samples from the Natural History Museum (NHM), London. Also many thanks to H.Fujiwara (NAO, Japan), A.Weinberger (DTM), C.Melis (UCSD) Y.K. Okamoto (Ibaraki Univ.) and C.Chen (STScI) for providing infrared spectra.

Part of this work (HD23514, HD15407a, HD165014, HD172555, BD+20307, ID8, HD145263, ID1211, HD113766, Eta Corvi and HD69830 ) is based on observations made with the Spitzer Space Telescope, which is operated by the Jet Propulsion Laboratory, California Institute of Technology. Another Part of this work (Beta Pictoris) is based on data collected at the SUBARU telescope, which is operated by the National Astronomical Observatory of Japan.

**Figure captions**

Figure 1a: Mid-infrared spectra of Martian samples (in µm and absorbance units). The heavily shocked shergottite Los Angeles is on top, followed by the other members of the shergottite group. The similar nakhlites follow, with chassignites at the bottom (for details about composition and sources see Table 2).

Figure 1b: Mid-infrared spectra of terrestrial samples (in µm and absorbance units). At the top are glasses - impact melts (tektites), obsidian and $SiO_2$. Samples of crustal material are granites, anorthosite, basalt and gabbro. Terrestrial mantle is represented by typical peridotites (dunite, harzburgite, orthopyroxenite). For details about composition and sources see Table 2.

Figure 2a. Comparison of laboratory spectra (grey) with astronomical spectra of debris disks from group A (black), which is characterized by features at < 9.5 µm and around 20 µm. At the top are representatives of heavily shocked crustal/upper mantle materials from differentiated bodies (tektites, mesosiderite, shocked shergottites), as well as obsidian and $SiO_2$. For details about the debris disks see Table 1.

Figure 2b. The suite of debris disk spectra of group B, compared with mafic, mantle-type materials (harzburgite, chassignite and brachinite). Here we see a similarity with the strong olivine bands at 10 and 11 µm, and features above 20 µm. This indicates two distinguishable groups of planetary dust materials from differentiated, potentially terrestrial planets. For details about the debris disks see Table 1.

Figure 3a-c. Continuum divided, base lined and normalized mid-infrared spectra of selected debris disks compared to base lined and normalized laboratory data. Bands A (9.0 - 9.6 µm) and B (10.8 and

11.4 µm) are the integrated areas which are used for ratios in Fig.4. (a) Dust spectra of group A show the dominance of the strongest band from ~9.0 µm to 9.5 µm with HD23514, HD172555 and HD165014 as examples for debris disks. Terrestrial impact melt glasses (tektite) and heavily shocked Martian shergottite Los Angeles 001 are used as comparable planetary materials. (b) and (c): In group B the main feature shifts from ~10 µm to ~11 µm, with the intensity between bands A and B more even. Here terrestrial harzburgite and achondritic brachinites are used as analogs for dust in debris disks with increased olivine contents such as ID8. For systems dominated by olivine (e.g. the inner system of Beta Pictoris, Eta Corvi or HD113766), terrestrial dunites and achondritic ureilites provide comparable spectra.

Figure 4. Band position of the strongest mid-infrared band plotted against the ratios of bands A (integrated area between 9.0 and 9.6 µm) and B (area between 10.8 and 11.4 µm) allow the comparison between the laboratory and typical astronomical spectra of debris disks. Group B is divided into an intermediate subgroup where bands A and B are of comparable intensity and an olivine-rich subgroup where band B is strongest. The dotted line is the dividing line between spectral groups A (crustal or upper mantle-type material) and B (mantle-type materials).

| Star | Type | Age (Ma) | M(dust) in g | M(star) in ¤ | Dist. AU | Mineralogy (Lit.) | Ref |
|---|---|---|---|---|---|---|---|
| **HD23514** | F6V | 100-120 | $10^{26}$ | | 0.25-1 | qz, fsp | [1] [2] |
| **HD15407A** | F3/F5 V | 80-2100 | $10^{26}$ | | 0.6-1/0.4-19.2 | am.pyx ± qz ± ann.silica ± silica | [3] [4] [5] |
| **HD172555** | A5V | 12 | $10^{26}$ | 2 | 5.8 | am.silica, am/xt.ol, xt.pyx, sulph. | [6] |
| **HD165014** | A0V/F2V | | | | 0.7-4.4 | am/xt.pyx, minor xt.ol | [7] |
| **BD+20307** | F | 0.3-~1Gyr | $10^{24}$ | 2.2 | 0.85/0.2-18.2 | am/xt ol,pyx, am/xt. pyx, | [5] [8] [9] |
| **Beta Pictoris** | A5V | 12-20 | $10^{20}$(lump) | 1.8 | 0.2-2000 | am/xt. ol | [10], [11], [12], [13] |
| **ID8** | G6/7 | 35-38 | $10^{26}$ | | 0.4-~1 | am/xt.ol, am.pyx | [2] [5] [14] [15] |
| **Eta Corvi** | F2V | 1.4 Gyr | $10^{24}$ (warm) $10^{25}$ (cold) | 1.4 | 3 | am/xt pyx, xt ol, ±sulph ±silica | [14] [16] |
| **HD145263** | F0V | <10 | | | 1.8 | am/xt.ol, silica | [17] [18] |
| **P1121** | F9 IV | 100 | 'like BP' | | | | [4] [15] |
| **HD113766A** | F3/F5V | 10-16 | $10^{26}$ | | 1.8 / 0.6-11.8 | am/xt.ol, am/xt.pyx, ±sulph, ±phyllo | [5] [16] [19] |
| **HD69830** | K0V | 3-10 Gyr | $10^{21}$ | 0.86 | 0.93-1.16 / 0.7-1.5 | am/xt ol, am/xt.pyx, ± carbonate ±silica | [5] [20] |

**Table 1:** Characteristics of circumstellar and debris disks discussed in the text. Qz=quarz, pyx=pyroxene, ol=olivine, xt=crystalline, am=amorphous, sulph=sulphides, phyllo=phyllosilicates, fsp=feldspar. References: [1] Rhee et al., 2008 [2] Meng et al., 2012 [3] Fujiwara et al., 2012 [4] Melis et al.,2010 [5] Olofsson et al., 2012 [6] Lisse et al., 2009 [7] Fujiwara et al., 2010 [8] Song et al., 2005 [9] Weinberger et al., 2011 [10] Chen et al., 2007 [11] Okamoto et al., 2004 [12] Telesco et al., 2005 [13] Li et al., 2012 [14] Lisse et al., 2012 [15] Gorlova et al., 2004 [16] Chen et al., 2006 [17] Smith et al., 2008 [18] Honda et al., 2004 [19] Lisse et al., 2008 [20] Lisse et al., 2007.

| Source | Main Mineralogy | Location/Specimen | Number |
|---|---|---|---|
| **Mars** | | | |
| Chassignite | **Olivine,** Ca-Pyroxene, Feldspars, Chromite | Chassigny | BM1997.2 |
| Shergottite | **Basaltic/Pyroxene-Phyric: Clinoyroxene (Augite, Pigeonite),** | Shergotty | BM41021 |
| | **Plagioclase/Maskelynite,** | Los Angeles | BM2000.M12 |
| | Fayalite, Ca-Phosphate, Ilmenite, Opaques | Zagami | BM1993.M10 |
| | **Olivine-Phyric: Olivine, Clinopyroxene (Pigeonite), Plagioclase/Maskelynite,** | SAU 005[1] | BM2000.M49 |
| | Opaques, Phosphate | DaG 476[2] | NHM, unreg. |
| Nakhlite | **Ca-Pyroxene (Augite), Olivine,** | Nakhla | BM1913.25 |
| | Plagioclase, oxides, sulfides, Silicate glass | MIL 03346[3] | NHM, unreg. |
| | | Governador Valadares | BM1975.M16 |
| | | Y000593[4] | NHM, unreg |
| | | Lafayette | BM1959.755 |
| **Earth** | | | |
| **Crust** | | | |
| Granite | **20-60% Quartz, 10-65%Plagioclase, Alkali-Feldspar,** Biotite, Muscovite | India | NHM unreg. |
| Gabbro | **10-90%Plagioclase, Pyroxene,** FeTi-oxides, Olivine, Hornblende, Biotite | India | NHM unreg. |
| Anorthosite | >90%Plagioclase, Olivine, Pyroxene | Niger | BM.1980,P8(1) |
| Basalt | **Plagioclase, Pyroxene, FeTi-oxides,** olivine | Iceland, Krafla | BM.1982,P4(38) |
| | contains glass | Canary Islands | BM.1973,P11(13) |
| | | India | NHM unreg. |
| **Mantle** | | | |
| Dunite | >90% Olivine, Pyroxene | New Zealand, South Island | BM.1974,P43(1) |
| Harzburgite | **40-90% Olivine, <10% Orthopyroxene** | New Zealand, South Island | BM.1973,P22(3) |
| Pyroxenite/Websterite | **Orthopyroxene, Clinopyroxene** | Pakistan, Karakoram | BM.1985,P29(25) |
| **Impact Rocks** | | | |
| Tektites | **Silicate Glass** | Australite Bediasite | NHM unreg. |

**Table 2:** Sources, petrology and mineralogy of bulk samples used in this study. Bold=dominating phases. NHM=Collection of the Natural history Museum, London, Literature: (Terrestrial: Best and Christiansen, 2001; Martian: McSween and Treiman, 1998).[1]Sayh al Uhaymir 005 [2]Dar al Gani 476 [3]Miller Range 03346 [4]Yamato 000593.

| Mass (g) | System | Source |
|---|---|---|
| $3.4 \times 10^{26}$ | HD113766A | [2] |
| $3.0 \times 10^{26}$ | HD15407A | [11] |
| $3.0 \times 10^{26}$ | ID8 | [4] |
| $1.0 \times 10^{26}$ | HD172555 | [3] |
| $1.0 \times 10^{26}$ | HD23514 | [7] |
| $6.0 \times 10^{25}$ | Eta Corvi (cold) | [4] |
| $9.0 \times 10^{24}$ | Eta Corvi (warm) | [4] |
| $2.3 \times 10^{24}$ | BD+20 307 | [6] |
| $2.0 \times 10^{21}$ | HD69830 | [1] |
| $4.0 \times 10^{20}$ | Beta Pictoris ('Lump') | [8] |
| $6.0 \times 10^{27}$ | Earth | [4] |
| $4.0 \times 10^{27}$ | Earth Mantle | [5] |
| $2.4 \times 10^{25}$ | Earth Crust | [5] |
| $1.5 \times 10^{25}$ | Earth Continental Crust | [5] |
| $1.0 \times 10^{25}$ | Earth Granite | [9] |
| $6.0 \times 10^{26}$ | Mars | [4] |
| $2.7 \times 10^{25}$ | Escaping Mass Mars Impact | [10] |
| $3.6 \times 10^{23}$ | Orbiting Mass Mars Impact | [10] |
| $4.9 \times 10^{27}$ | Venus | [5] |
| $3.3 \times 10^{26}$ | Mercury | [3] |
| $7.0 \times 10^{25}$ | Moon | [4] |
| $1.0 \times 10^{25}$ | Pluto | [4] |
| $3.0 \times 10^{24}$ | Asteroid Belt | [4] |
| $1.0 \times 10^{24}$ | Asteroids (maximal) | [4] |
| $1.0 \times 10^{16}$ | Asteroids (minimal) | [4] |

**Table 3.** Dust masses for various debris disk systems in comparison to the masses of planetary structures and bodies (in g). References: [1]= Lisse et al., 2007 [2] Lisse et al., 2008 [3] Lisse et al., 2009 [4] Lisse et al., 2012 [5] Lodders and Fegley Jr., 1998 [6] Olofsson et al., 2012 [7] Rhee et al., 2007 [8] Telesco et al., 2005 [9] Bonin 2012 [10] Marinova et al., 2011 [11] Fujiwara et al., 2010.

| **Shergotittes** | Pyx | Plag Pyx | Plag | Ol Pyx | Pyx | Pyx | Pyx | Pyx | Plag Pyx | Pyx | Pyx |
|---|---|---|---|---|---|---|---|---|---|---|---|
| **Los Angeles** | 9.52 | 10.31 | | 11.2 (sh) | 13.74 | 14.9 | 15.87 | | 20.96 | | 30.5 |
| **Zagami** | 9.38 | 10.35 | 10.74 | 11.33 | 13.68 | 14.88 | 15.8 | 19.88 | | 25.84 | 30.21 |
| **Shergotty** | 9.41 | 10.36 | | 11.31 | 13.68 | 14.9 | 15.8 | 19.88 | | 25.9 | 30.21 |
| **DaG476** | 9.47 | 10.38 | 10.6 | 11.39 | 13.66 | 14.84 | 15.67 | 19.88 | | 25.06 | 29.5 |
| **SaU005** | 9.47 | 10.34 | 10.61 | 11.33 | 13.66 | 14.84 | 15.72 | 19.88 | | 24.81 | 29.15 |
| **Average** | **9.45** | **10.35** | **10.65** | **11.34** | **13.68** | **14.87** | **15.77** | **19.88** | **20.96** | **25.40** | **29.91** |
| **w/o Los Angeles** | **9.43** | **10.36** | **10.65** | **11.34** | **13.67** | **14.87** | **15.75** | **19.88** | | **25.40** | **29.77** |

| **Nakhlites** | Pyx | Plag Pyx | Ol Pyx | | Pyx | Pyx | Pyx | Plag Pyx | Pyx | Pyx | Pyx |
|---|---|---|---|---|---|---|---|---|---|---|---|
| **Lafayette** | 9.37 | 10.33 | 11.4 | | 14.88 | 15.8 | 19.88 | 20.88 | 25.84 | 30.21 | 33.5 |
| **Governador** | 9.36 | 10.31 | 11.39 | | 14.86 | 15.8 | 19.88 | 20.83 | 25.84 | 30.21 | 33.5 |
| **Y000593** | 9.36 | 10.31 | 11.4 | | 14.84 | 15.82 | 19.88 | 20.88 | 25.84 | 30.12 | 33.5 |
| **Nakhla** | 9.36 | 10.32 | 11.39 | | | 15.82 | 19.88 | 20.88 | 25.84 | 30.21 | 33.5 |
| **MIL03346** | 9.36 | 10.3 | 11.39 | | 14.84 | 15.82 | 19.84 | 20.88 | 25.58 | 30 | 33.5 |
| **Average** | **9.36** | **10.31** | **11.39** | | **14.86** | **15.81** | **19.87** | **20.87** | **25.79** | **30.15** | **33.5** |

| **Chassignites** | | Ol | Ol | Ol | Ol | Ol | Ol | Ol | Ol | Ol | Ol |
|---|---|---|---|---|---|---|---|---|---|---|---|
| **Chassigny** | | 10.29 | 11.33 | 11.98 | 16.92 | 19.42 | 20.24 | 21.74 | 24.81 | 28.5 | 35 |

**Table 4a.** Positions of probable main infrared features from laboratory spectra of Martian, terrestrial and asteroidal samples (in µm). Sh=Shoulder. Best candidates for phases responsible for feature: Pyx=Ca-rich pyroxene, Ol=Olivine, Plag=Plagioclase, Kfsp=Alkalifeldspar, Qtz=Quarz, Sig=Silica Glass. For details see Table 5.

**Earth**

| Tektites/Glass | Sig | | | | | | Sig | | | | | Sig | | |
|---|---|---|---|---|---|---|---|---|---|---|---|---|---|---|
| SiO₂ | 8.97 | | | | | | 12.17 | | | | | | | |
| Bediasite | 9.17 | | | | | | 12.55 | | | | | 21.63 | | |
| Australite | 9.3 | | | | | | 12.63 | | | | | 21.69 | | |
| Obsidian | 9.3 | | | | | | 12.2 | | | | | | | |

| Crustal | K-Fsp Plag | Qz Plag | Plag | Plag | Plag | Pyx | Qz Kfsp | Plag | Kfsp Plag | Kfsp Plag | Kfsp Plag | Qz Plag | Kfsp Plag | |
|---|---|---|---|---|---|---|---|---|---|---|---|---|---|---|
| Granite A | 8.77 | 9.15 | 9.65 | 9.85 | 9.95 | | 12.84-13.77 | | 15.43 | 16.95 | 18.77 | 21.55 | 23.36 | |
| Granite B | 8.77 | 9.15 | 9.63 | 9.85 | | | 12.86-14.39 | | 15.43 | 17.03 | 18.76 | 21.55 | 23.35 | |
| Anorthosite | 8.7(sh) | 9.16 | | 10 | 10.6sh | | | 13.5 | | 17.21 | 18.48 | 21.46 | | |
| Basalt A | 8.7(sh) | 9.2(sh) | 9.7(sh) | 10.09 | | | | 12.83-13.8 | 15.53 | 16.98 | | 21.79 | 23.14 | |
| Basalt C | | 9.14 | 9.79 | | | | | 13.9 | 15.82 | 17.22 | 18.43 19.34 | 21.23 | 23.35 | |
| Gabbro | 8.7(sh) | 9.25 | | 10.04 | | 11.4 | | 12.53-12.83 | 15.85 | 17.01 | | 21.76 | | |

| Mantle | | Pyx | | Ol Pyx | Ol Pyx | Ol Pyx | Ol | Pyx | Ol Pyx | Ol l | Ol | Ol Pyx | Ol Pyx | Ol Pyx | Ol Pyx |
|---|---|---|---|---|---|---|---|---|---|---|---|---|---|---|---|
| Dunite | | | | 10.08 | 10.46 | 11.3 | 11.93 | | 16.45 | 19.8 | 21.45 | 23.98 | 19.8 | 27.7 | 33.9 |
| Harzburgite | | 9.3(sh) | | 10.17 | 10.48 | 11.29 | 11.91 | | 16.37 | 19.8 | | 23.81 | 19.8 | 27.55 | 33.78 |
| Orthopyroxenite | | 9.31 | | 10.17 | 10.86 | 11.42 | | 15.75 | | 19.46 | 21.93 | | 19.46 24.88 | 29.59 | 34.25 |

| Diff.Asteroids | | Pyx Qz | Pyx | Ol | Ol | Ol | Ol | | Pyx | Ol | Ol | Ol | Ol | Ol | Pyx Qz |
|---|---|---|---|---|---|---|---|---|---|---|---|---|---|---|---|
| Emery | | 9.11 | 10.33-10.54 | | | | | 12.59 | 15.43 | | | | | 21.17 | |
| Brachina | | | | 10.11 | 10.58 | 11.27 | 11.95 | 12.5 | | 16.89 | 19.35 | 19.94 | 21.79 23.9 | | |

**Table 4b.**

| Mineral | | | | | | | | | | | |
|---|---|---|---|---|---|---|---|---|---|---|---|
| **Olivine (Ol)** | 10.0-10.6 | 10.4-10.9 | 11.2-11.4 | 11.9-12.1 | 16.4-17.7 | 18.2-19.6 | 19.7–20.9 | 23.7-27.7 | 27.5–31.9 | 33.8–38.9 | 19.7–20.9 |
| **Ca-rich Pyroxene (Pyx)** | 9.3 | 10.3 | 11.4 | 15.7-15.8 | 14.82 | 19.3-19.5 | 20.6 | 24.3-25.1 | 29.5-29.6 | 32.1-32.3 | 33.7-33.9 |
| **Carbonate (Car)** | 7 | ~11 | ~14 | | | | | | | | |
| **Plagioclase** | 8.6-8.7 | 9.1 | 9.6-10.1 | 10.5-10.8 | 12.7-13.8 | 14.9-15.4 | 16.1-16.4 | 16.9-17.3 | 18.5-18.8 | 20.6-21.6 | 23.5 |
| **Alkali Feldspar** | 8.8 | 9.7-9.9 | 13-13.8 | 15.4-15.6 | 17.1 | 18.5-18.6 | 23.4-23.8 | | | | |
| **Quarz (Qz)** | 8.9-9.2 | 12.4-12.8 | 20.8-21.7 | | | | | | | | |
| **Silica Glass (Sig)** | 9.0 | 12.2 | 18.3 | | | | | | | | |

**Table 5:** Characteristic features for the mineral bands of the phases occurring in the analyzed rock (in µm). Olivine: ranges from Mg-rich end member Forsterite to Fe-rich end member Fayalite (Koike et al., 2003; Morlok et al., 2006). Pyroxenes are Ca-rich Diopside and Augite (Koike et al., 2000). Carbonates are for Ca-, Mg-, Fe- and Mn-Carbonates (Koike et al., 2000). Quartz (Qz) is from Farmer (1974) and Koike et al. (1994). Alkali feldspar is from orthoclase and microcline (ASTER Spectral Library Library; http://speclib.jpl.nasa.gov/documents/jhu_desc). Plagioclase members are albite, labradorite and anorthite (ASTER spectral library Library http://speclib.jpl.nasa.gov/documents/jhu_desc). Representative for silica rich glass is amorphous $SiO_2$, but changes in composition (e.g. Fe or Mg contents in tektites) can lead to significant shifts in band positions (Speck et al., 2011).

| Debris Disks | | | | | | | | | | | | | | | | | |
|---|---|---|---|---|---|---|---|---|---|---|---|---|---|---|---|---|---|
| **HD23514**   | 8.97 |       |       | 12.6  |       |       |         | 20.76 |       |       | 25.84 |       | 28.9 |      | 31.8 |      |      | 34.6  |
| **HD15407A**  | 9.09 |       |       | 12.5  | 13.99 |       | 15.93   | 20.08 |       |       | 24.32 |       |      |      |      |      |      | 33.3  |
| **HD172555**  | 9.32 | 9.44  | 11.1  | 12.4  |       |       |         | 19    |       |       |       |       |      |      |      |      |      |       |
| **HD165014**  | 9.33 | 10.42 | 11.08 |       |       | 14.66 |         | 18    |       |       |       |       | 27.9 |      |      |      |      | 34    |
| **BD+20307**  | 9.87 | 10.24 | 10.84 |       |       |       | 16.43   | 18.97 |       |       | 23.54 |       | 29.9 |      | 31.4 |      |      | 33.9  |
| **Beta Pic**  | 9.98 |       | 11.07 |       |       |       |         |       |       |       |       |       |      |      |      |      |      |       |
| **ID8**       | 10   |       | 11    |       |       |       | 16.8(sh)| 19.3  |       | 21.7  | 23    | 23.8  |      |      |      |      |      |       |
| **P1121**     | 10   |       | 11    |       |       |       | 16.6    | 18.5  |       | 21.8  | 23    | 24.1  | 27.2 |      |      | 28.1 | 32.8 |       |
| **HD113766**  | 9.98 |       | 11.12 |       | 14.04 | 14.9  | 16.24   | 18.92 | 19.67 | 21.01 | 23.1  | 23.7  |      | 25.1 | 27.5 | 29.2 | 31.6 | 34.3  |
| **HD69830**   | 10.3 |       | 10.93 | 11.96 | 13.6  |       | 16.67   | 18.96 |       |       |       | 23.95 |      |      |      |      |      | 33.43 |
| **Eta Corvi** |      |       | 11.3  | 11.9  |       |       | 16.6    |       | 19.5  |       |       |       |      |      |      |      |      |       |
| **HD145263**  | 9.88 |       | 11.15 |       | 14.05 |       |         |       | 19.6  |       |       |       |      |      |      |      |      | 33.3  |

Table 6: Main bands from astronomical spectra of debris disks (in µm). Sh=Shoulder.

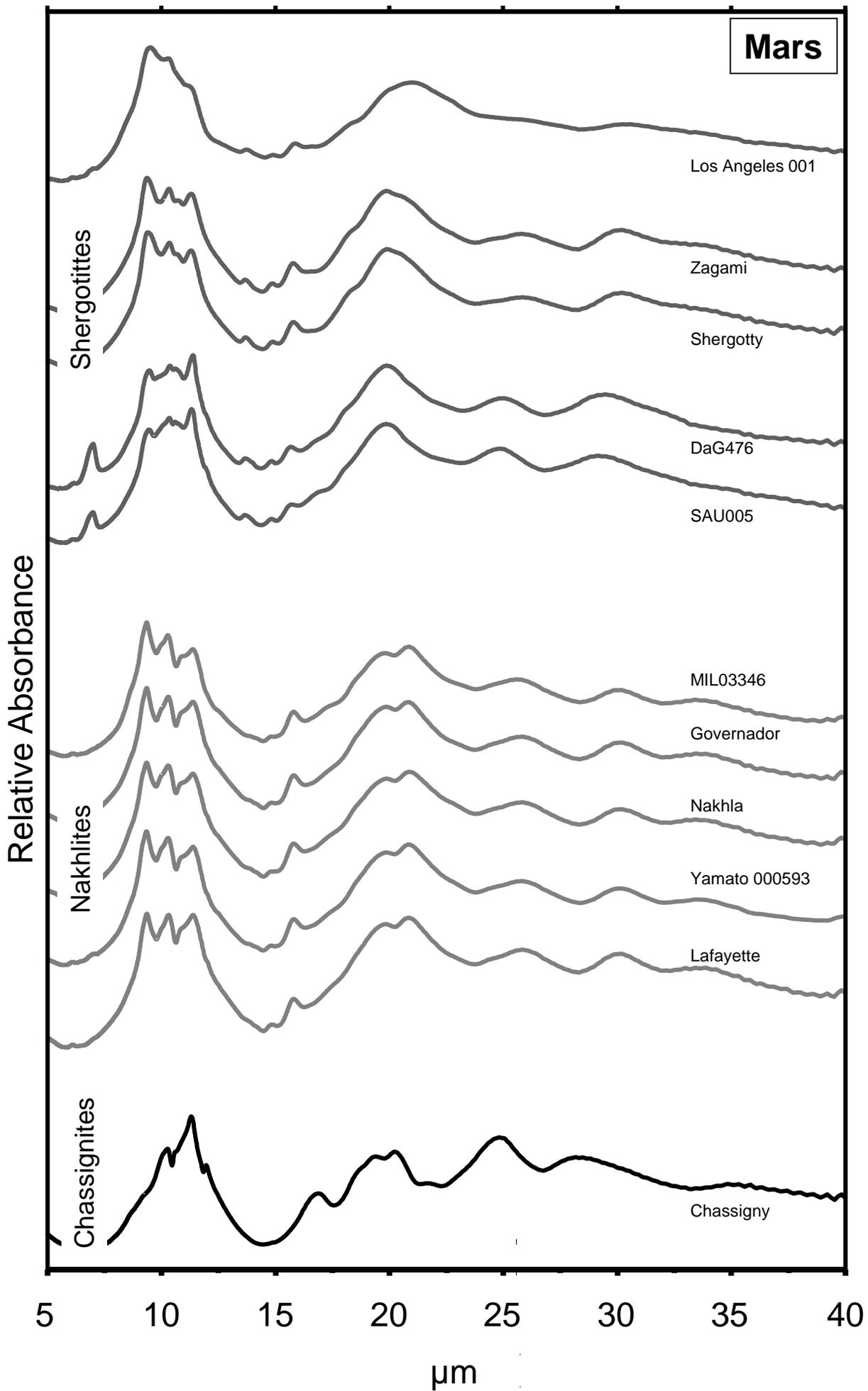

Figure 1a

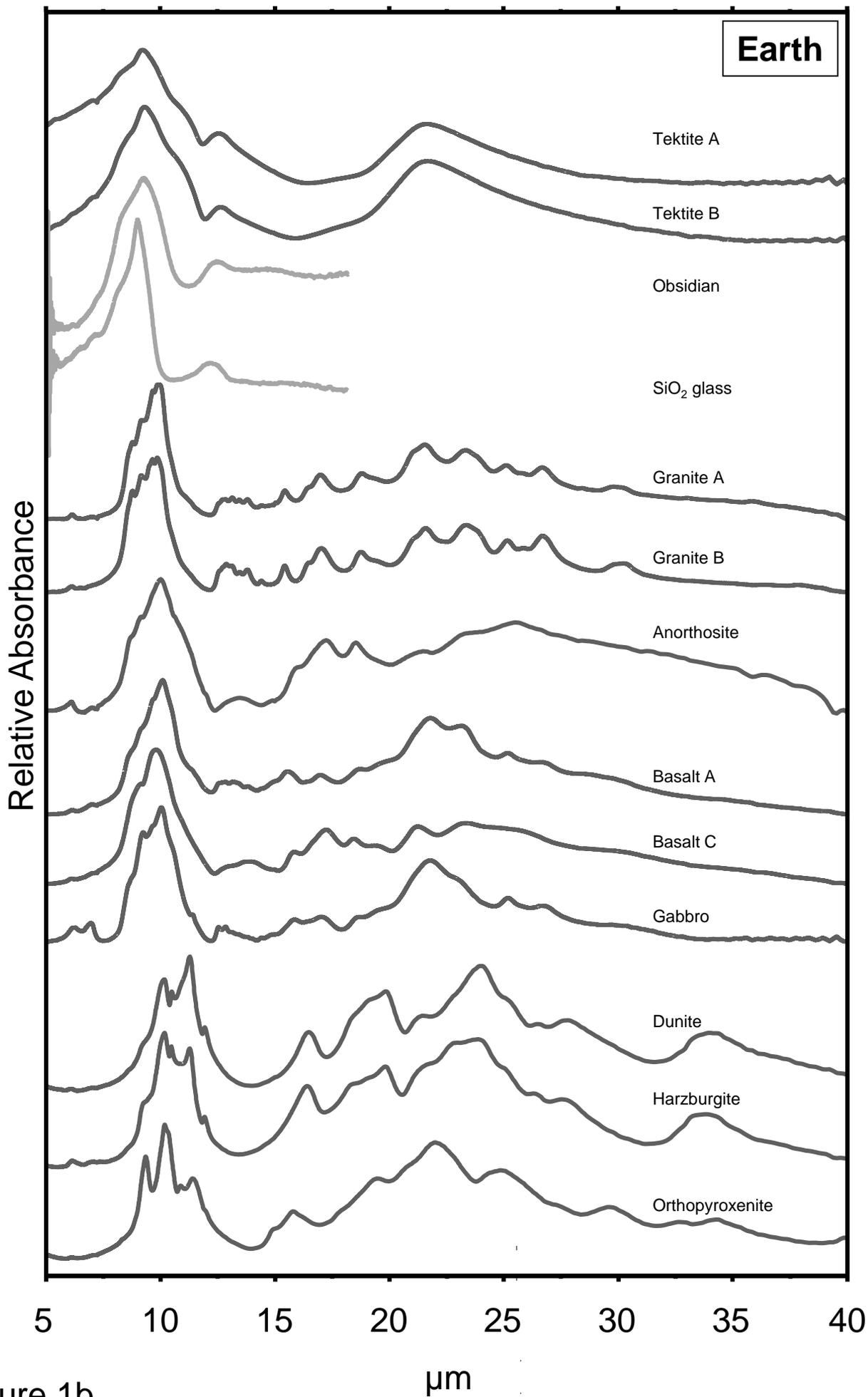

Figure 1b

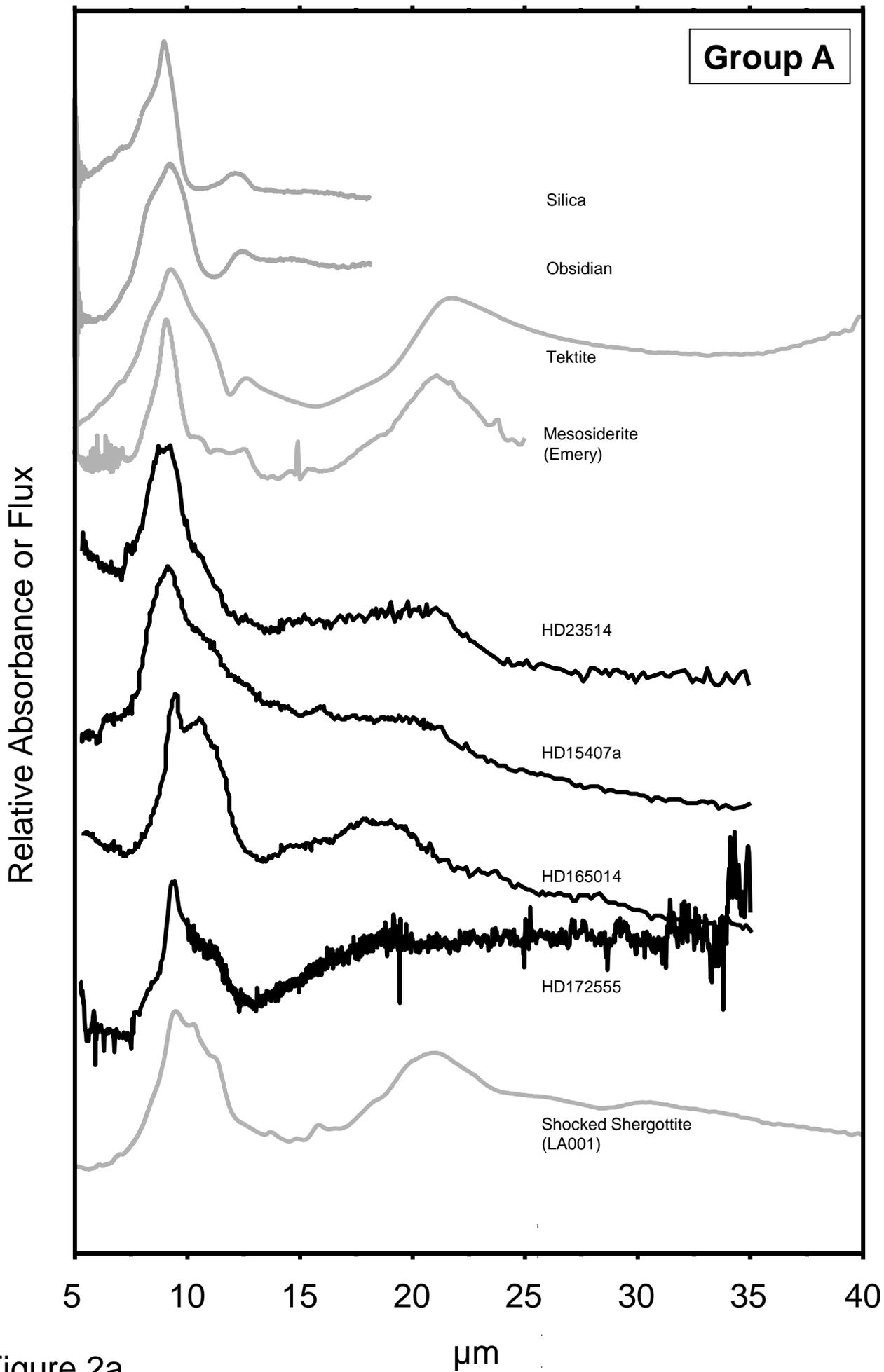

Figure 2a

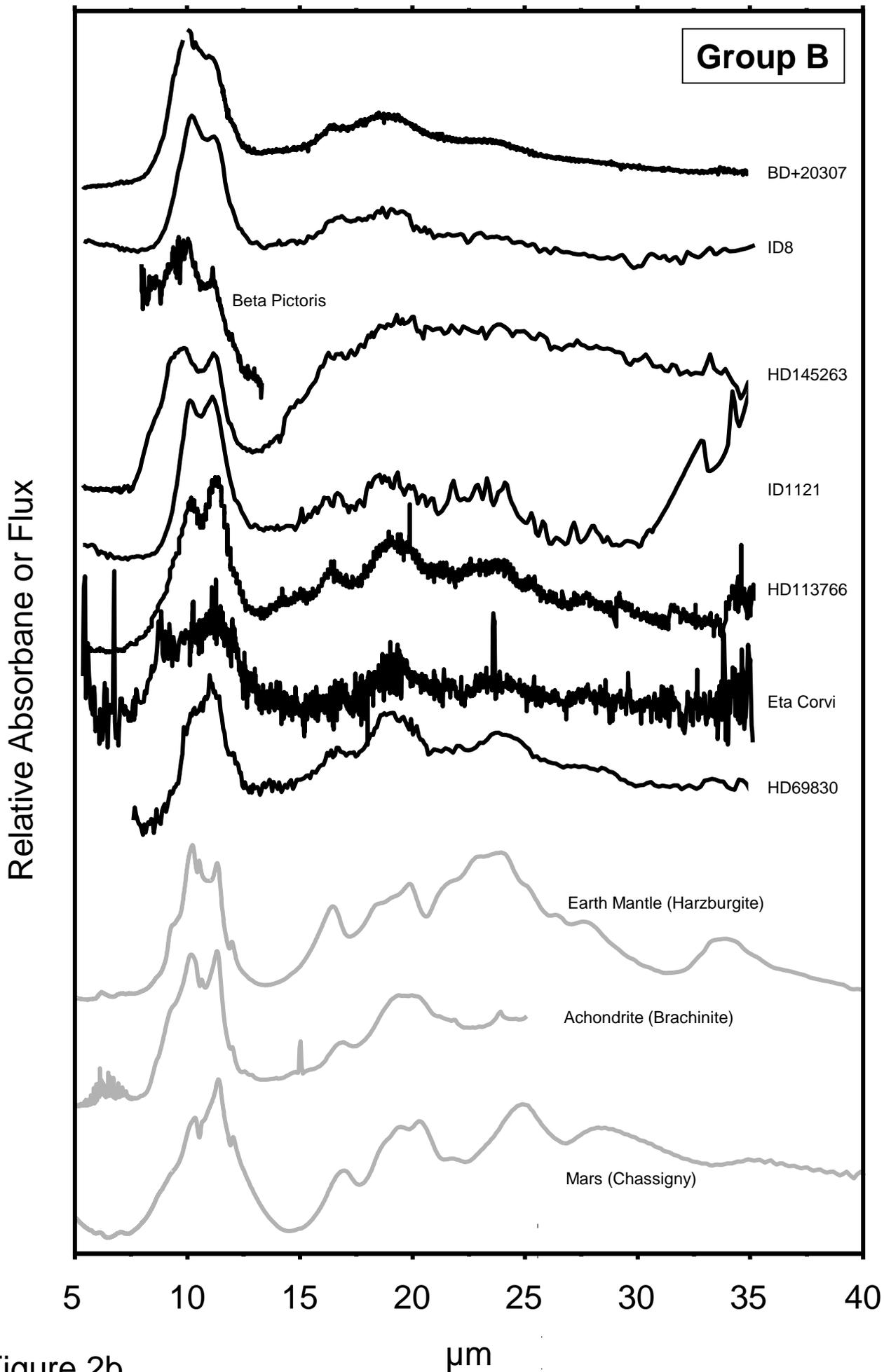

Figure 2b

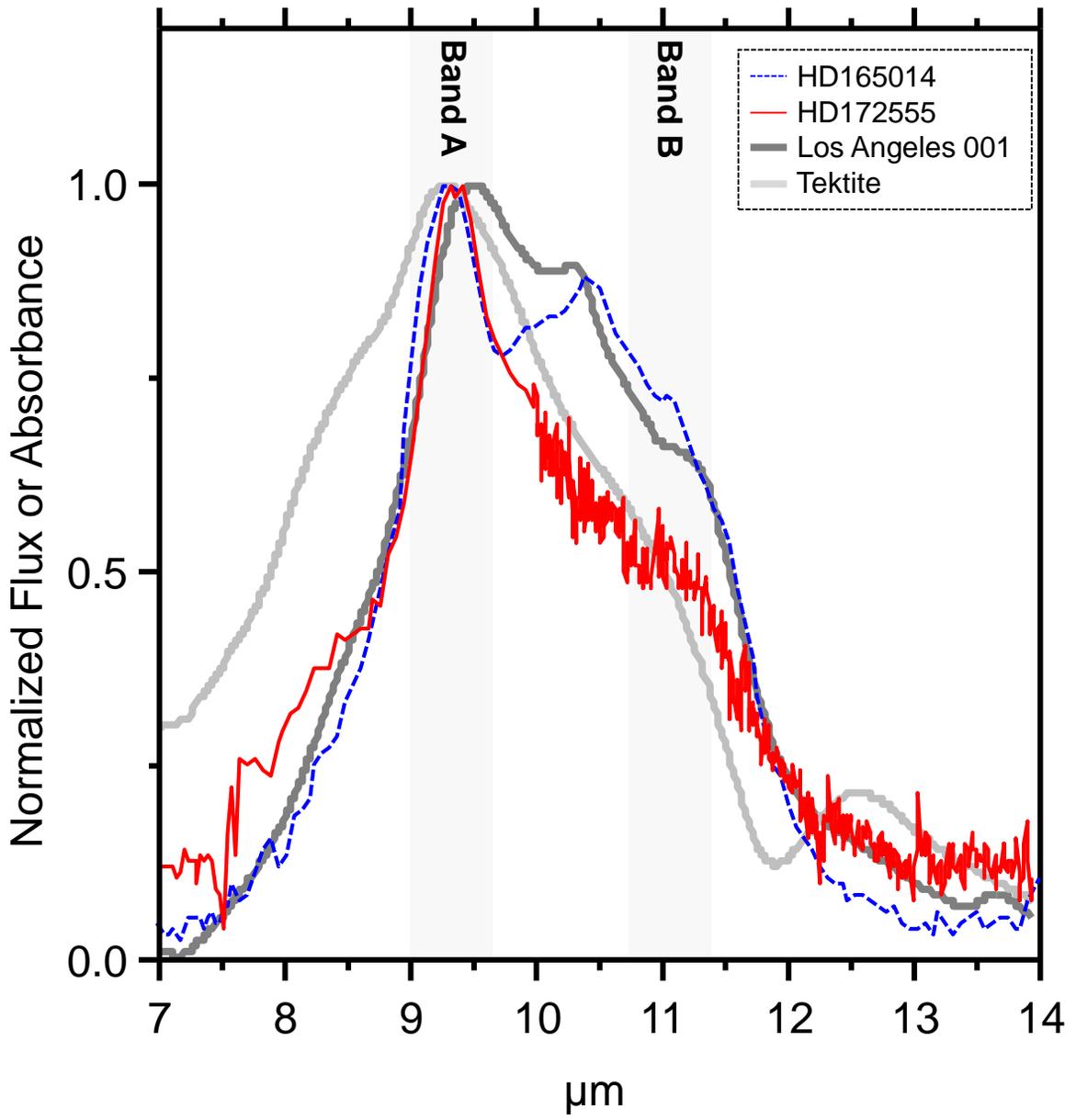

Figure 3a

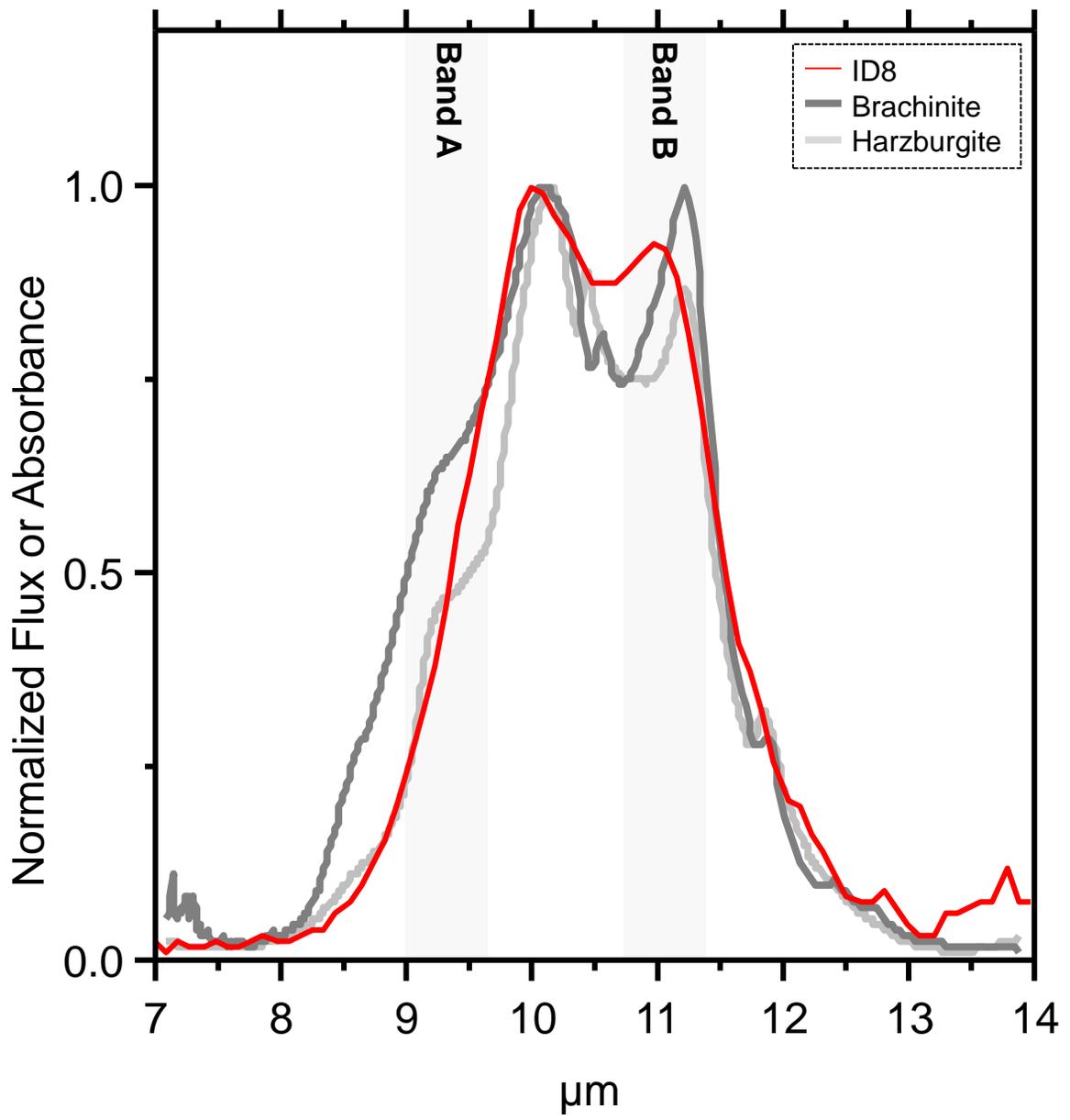

Figure 3b

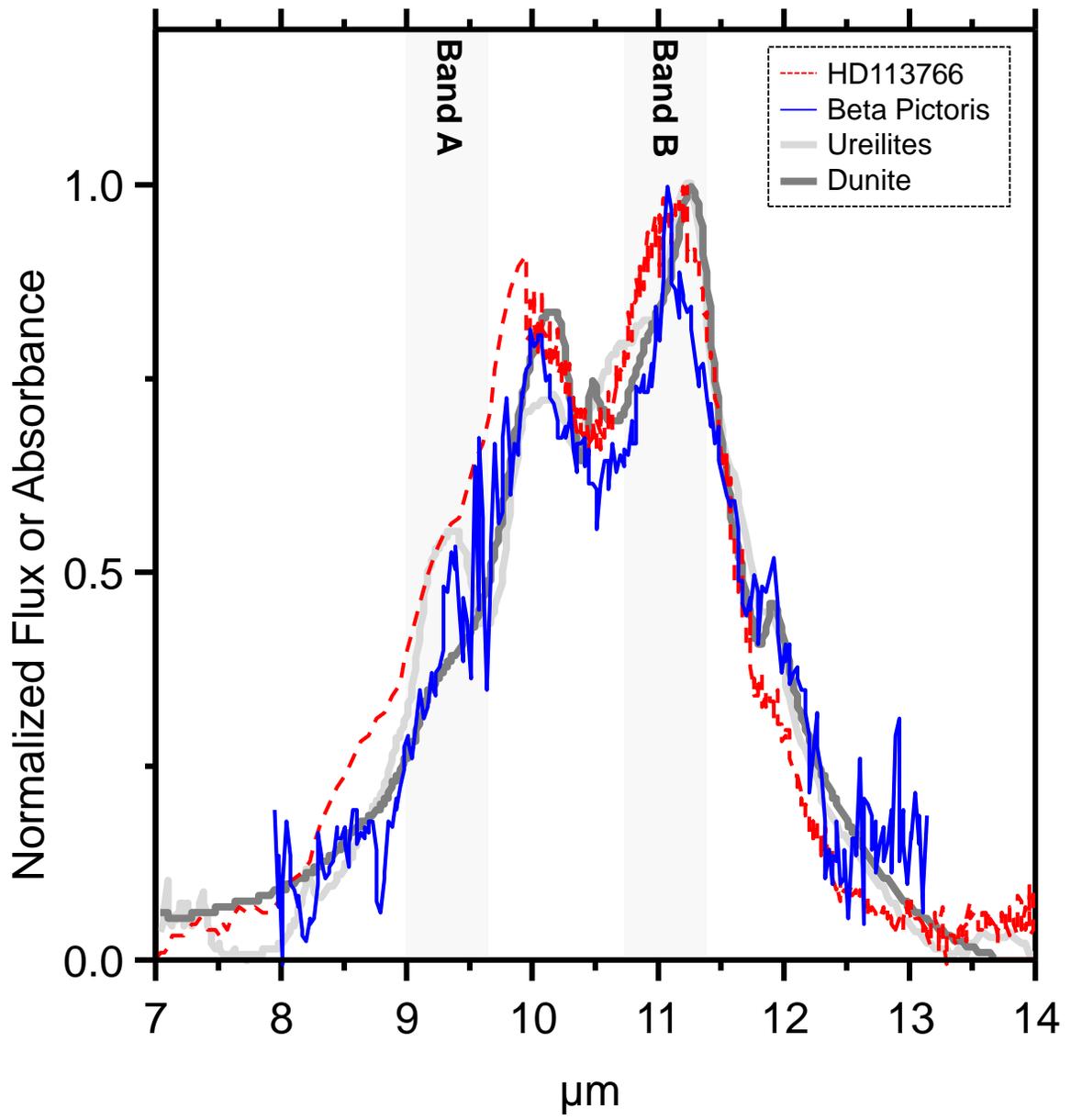

Figure 3c

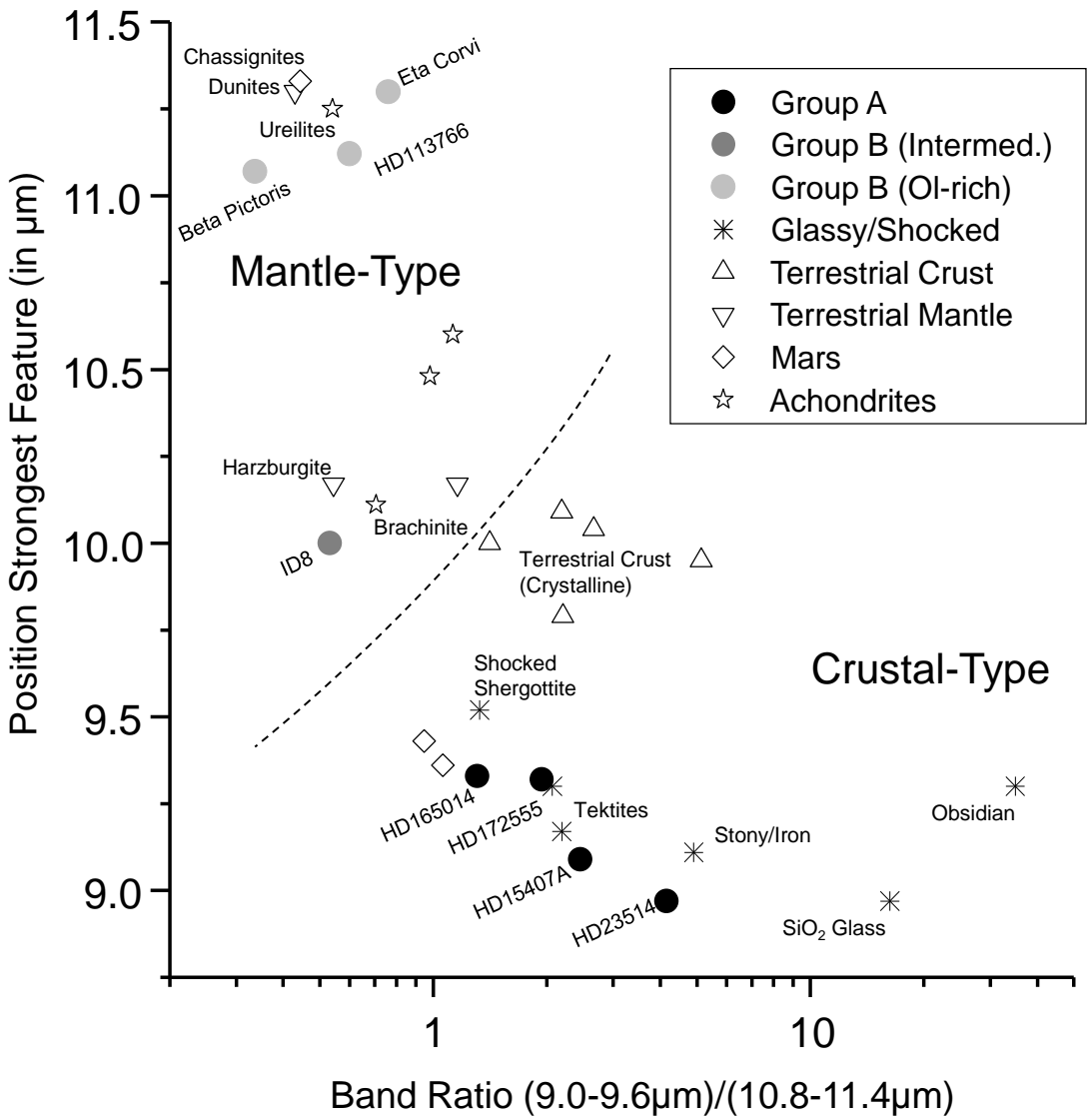

Figure 4